\documentclass[aps,twocolumn,showpacs,preprintnumbers,prl,superscriptaddress,groupedaddress,10pt]{revtex4-1}
\usepackage[utf8]{inputenc}
\usepackage{graphicx,amssymb,amsmath,amsthm,amsfonts,epsfig,times,natbib}

\usepackage[usenames,dvipsnames]{color}
\usepackage{epstopdf}
\usepackage{amsmath,amssymb}
\usepackage{tensor}
\usepackage{mathtools}
\usepackage{amsbsy}
\usepackage{bm,url}
\usepackage[linktocpage]{hyperref}

\usepackage[scaled]{beramono}
\usepackage[T1]{fontenc}

\definecolor{oxfordblue}{rgb}{0.0, 0.13, 0.28}
\definecolor{burgundy}{rgb}{0.5, 0.0, 0.13}
\definecolor{darkolivegreen}{rgb}{0.33, 0.42, 0.18}
\definecolor{darkblue}{rgb}{0,0,0.5}
\definecolor{richcarmine}{rgb}{0.84, 0.0, 0.25}
\definecolor{darkblue}{rgb}{0,0,0.5}
\definecolor{venetianred}{rgb}{0.78, 0.03, 0.08}
\definecolor{skobeloff}{rgb}{0.0, 0.48, 0.45}
\hypersetup{colorlinks=true, citecolor=darkblue, linkcolor=darkblue,
urlcolor = darkblue}

\newcommand{\ben}{\begin{enumerate}}
\newcommand{\een}{\end{enumerate}}

\def\be{\begin{equation}}
\def\ee{\end{equation}}
\def\bea{\begin{eqnarray}}
\def\eea{\end{eqnarray}}

\newcommand{\beq}{\begin{eqnarray}}
\newcommand{\eeq}{\end{eqnarray}} 
\newcommand{\ba}{\begin{align}}
\newcommand{\ea}{\end{align}}

\begin{document}

\title{Stationary black holes and light rings}

\author{
Pedro V. P. Cunha$^{1}$,
Carlos A. R. Herdeiro$^{2}$,
}

\affiliation{${^1}$ Max Planck for Gravitational Physics - Albert Einstein Institute, Am M\"uhlenberg 1, Potsdam 14476, Germany }
 \affiliation{${^2}$ Departamento de Matem\'atica da Universidade de Aveiro and CIDMA,
Campus de Santiago, 3810-183 Aveiro, Portugal}
%


\date{March 2020}

\begin{abstract}
The ringdown and shadow of the astrophysically significant Kerr Black Hole (BH) are both intimately connected to a special set of bound null orbits known as Light Rings (LRs).  Does it hold that a \textit{generic} equilibrium BH \textit{must} possess such orbits? In this letter we prove the following theorem. A stationary, axi-symmetric, asymptotically flat black hole spacetime in 1+3 dimensions, with a non-extremal, topologically spherical, Killing horizon admits, at least, one standard LR outside the horizon for each rotation sense. The proof relies on a topological argument and assumes $C^2$-smoothness and circularity, but makes no use of the field equations. 
The argument is also adapted to recover a previous theorem establishing that a horizonless ultra-compact object must admit an even number of non-degenerate LRs, one of which is stable. 
\end{abstract}




\maketitle
{\bf Introduction.}
The second decade of the XXIst century will be celebrated as the dawn of precision strong gravity. New observational data is testing, in particular, the true nature of astrophysical black holes (BHs). Both gravitational wave observations~\cite{Abbott:2016blz,LIGOScientific:2018mvr}, notably through the ringdown phase, and electromagnetic observations, in particular through the shadow imaging~\cite{Akiyama:2019cqa,Akiyama:2019fyp,Akiyama:2019eap}, are expected to provide hitherto inaccessible information on the BH spacetime geometry. 

The ringdown and shadow observables are both intimately connected to a special set of bound null orbits for test particles~\cite{Cardoso:2016rao,Cunha:2018acu}. When planar, these orbits are known as \textit{light rings} (LRs). They are an extreme form of light deflection, such that the path of light closes over itself.  In the general non-planar case these light paths are dubbed \textit{fundamental photon orbits} (FPOs)~\cite{Cunha:2017eoe}. For a spherical BH, such as the Schwarzschild solution, all FPOs are LRs. This is not so for axi-symmetric, but non spherical, BHs. In the special case of the Kerr spacetime,  the FPOs are known as \textit{spherical photon orbits}~\cite{Teo2003}, all of which are unstable (in the radial direction) outside the horizon and reduce, in two appropriate limits, to LRs. The latter correspond to equatorial photon orbits which are co-rotating/counter-rotating with the Kerr horizon. 

The close connection between LRs and the aforementioned key observables raises the following question: does an equilibrium BH spacetime always possesses LRs? This is the case for the paradigmatic electro-vacuum BHs of General Relativity (GR), but can one safely extrapolate to BHs with generic matter contents or modified gravity?

 In this letter we shall provide a generic and robust answer to these questions using a topological argument. Concretely, under reasonable assumptions, we shall establish the following theorem: \textit{a stationary, axi-symmetric, asymptotically flat, 1+3 dimensional BH spacetime, $(\mathcal{M},g)_{\rm BH}$, with a  non-extremal, topologically spherical Killing horizon, $\mathcal{H}$, admits at least one standard LR outside the horizon for each rotation sense.}

{\bf The spacetime.}
We assume an equilibrium BH spacetime under the conditions of the last paragraph. No assumption is made on the field equations  $(\mathcal{M},g)_{\rm BH}$ solves. This spacetime possesses two Killing vectors $\{\xi,\eta\}$, associated, respectively, to stationarity and axi-symmetry. Asymptotic flatness implies $\{\xi,\eta\}$ must commute~\cite{Carter:1970ea}. Then, coordinates ($t,\varphi$) adapted to the Killing vectors $\xi=\partial_t,\eta=\partial_\varphi$ can be chosen. In addition, we assume that the metric is at least $C^2$-smooth on and outside $\mathcal{H}$, and circular. The latter, together with asymptotic flatness, implies the spacetime admits a 2-space orthogonal to $\{\partial_t,\partial_\varphi\}$ - see, $e.g.$ theorem 7.11 in~\cite{Wald:1984rg}.  This means  the metric $g$ possesses a discrete symmetry  $(t,\varphi)\to (-t,-\varphi)$~\footnote{We do not assume a $\mathbb{Z}_2$ north-south spacetime symmetry.}.

In the orthogonal 2-space one can introduce spherical-like coordinates ($r,\theta$). The sections of $\mathcal{H}$  are assumed  to be topologically spherical. A  gauge choice guarantees the horizon is located at a constant (positive) radial coordinate $r=r_H$. The polar coordinate $\theta$ is chosen to be always orthogonal to $r$. In such a gauge, $g_{r\theta}=0$, $g_{rr}>0$ and $g_{\theta\theta}>0$ outside $\mathcal{H}$. One can further require that $(r,\theta)$ reduce to standard spherical coordinates in the asymptotically flat limit $r\to \infty$. The coordinates range is then, outside the horizon, $r\in[r_H,\infty[$, $\theta\in [0,\pi]$ with $\theta=\{0,\pi\}$ at the rotation axis, $\varphi\in[0,2\pi[$ and $t\in]-\infty,+\infty[$. Outside $\mathcal{H}$, causality requires $g_{\varphi\varphi}\geqslant 0$.  The metric, which has a Lorenzian signature $(-,+,+,+)$, thus reads  $ds^2=g_{tt}dt^2+2g_{t\varphi}dtd\varphi +g_{\varphi\varphi}d\varphi^2+g_{rr}dr^2+g_{\theta\theta}d\theta^2$.

{\bf The Killing horizon.}
The existence of $\mathcal{H}$ means there is a Killing vector field, $\chi =\partial_t + \omega_{_H}\partial_\varphi$, ($\omega_{_H}={\rm const.}$) that is null on $\mathcal{H}$,  $\left.\left(\chi^\mu\,\chi_\mu\right)\right|_\mathcal{H}=0$. 
Then, $\chi$ is the horizon null generator. 
For stationary BHs, one can further introduce a (positive) constant quantity on $\mathcal{H}$, the \textit{surface gravity} $\kappa$, defined via the following relation computed at the horizon $\left.[ \nabla_\mu(\chi^2)=-2\kappa\,\chi_\mu ]\right|_\mathcal{H}$.
Taking $\mu\in\{t,\varphi\}$, one obtains $0=\left.\left(g_{\mu t}+g_{\mu\varphi}\,\omega_{_H}\right)\right|_\mathcal{H}$. This implies that $\omega_{_H}=-\left.(g_{t\varphi}/g_{\varphi\varphi})\right|_\mathcal{H}$, for the horizon angular velocity $\omega_{_H}$, and $\left. D\right|_\mathcal{H}=0$, where we have defined $D\equiv (g^2_{t\varphi}-g_{tt}g_{\varphi\varphi})$. Thus, $D$ vanishes on $\mathcal{H}$; in fact, it is positive outside the horizon and away from the axis~\footnote{Notice that $-D<0$ is the determinant of the $t$-$\varphi$ sector of the metric}.


{\bf LRs and a topological charge.}
For diagnosing the occurrence of LRs in $(\mathcal{M},g)_{\rm BH}$, one must consider the null geodesic flow. Following~\cite{Cunha:2017qtt}, LRs are identified by considering the effective potentials on the orthogonal 2-space, $H_\pm$:
\begin{equation}
H_\pm(r,\theta)\equiv \frac{-g_{t\varphi}\pm\sqrt{D}}{g_{\varphi\varphi}} \ .
\label{poth}
\end{equation}
LRs are critical points of $H_\pm$; a LR obeys either $\partial_\mu H_+=0$ or  $\partial_\mu H_-=0$ or both simultaneously ($e.g.$ for static spacetimes)~\cite{Cunha:2016bjh}. The $\pm$ sign is typically associated to the two possible rotation senses (see Appendix A). 

We can associate a topological charge to LRs. 
First, introduce a field ${\bf v}=(v_r,v_\theta)$ as a normalised gradient of $H_\pm$:
\begin{equation}
v_r\equiv \frac{\partial_rH_\pm}{\sqrt{g_{rr}}} \ ,\qquad v_\theta\equiv \frac{\partial_\theta H_\pm}{\sqrt{g_{\theta\theta}}}\ .
\label{vrvt}
\end{equation}
If follows that $\partial^\mu H_\pm\,\partial_\mu H_\pm=v_r^2+v_\theta^2\equiv v^2$. Hence, in terms of ${\bf v}$, a LR occurs  iff ${\bf v}=0 \Leftrightarrow v=0$. 

Second, define an angle $\Omega$ such that $v_r=v\cos\Omega$,  $v_\theta=v\sin\Omega$. Then, $\Omega$ together with the ``norm" $v$,  parameterises the \textit{auxiliary 2-space} spanned by ${\bf v}$, denoted $\mathcal{V}$.

Third, in the physical orthogonal 2-space $(r,\theta)$, consider a simple closed curve $C$, that is piece-wise smooth and positive oriented. Since $C$ is closed, the angle $\Omega$ after a full revolution must be the same, modulo $2\pi$. Hence, 
\be
\oint_Cd\Omega=2\pi w \ , \qquad w\in\mathbb{Z} \ .
\label{tc}
\ee
In the physical $(r,\theta)$ space  $w$ counts the winding number of ${\bf v}$ as $C$ is circulated in the positive sense. 
When $C$ encloses a single (non-degenerate~\cite{Cunha:2017qtt,Hod:2017zpi})\, LR, the integer $w$ is the \textit{topological charge of the LR}. Indeed, the curve $C$, in the physical $(r,\theta)$ space, defines a curve $\widetilde{C}$ in $\mathcal{V}$, via~\eqref{vrvt}. In $\mathcal{V}$,  $w$ is the winding number of $\widetilde{C}$ around the origin $(v=0)$, which corresponds to a LR. Thus, in  $\mathcal{V}$, $w$ constitutes a well defined topological quantity~\footnote{One can also regard $w$ as the Brouwer degree of the map defined by the angle $\Omega$, namely $\Omega: S^1 \to S^1$, which maps each point of $C$ to a point in an auxiliary circle. This was the perspective taken in~\cite{Cunha:2017qtt}.}: deforming $\widetilde{C}$ without crossing the origin does not change $w$. Consequently, in the physical $(r,\theta)$ space, deforming $C$ without crossing a LR does not change $w$. 

\begin{figure}[t!]
\begin{center}
\includegraphics[width=0.45\textwidth]{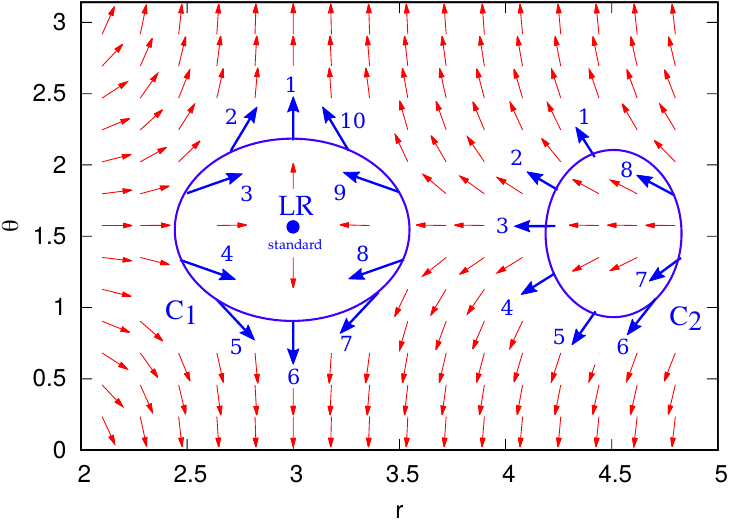}
\caption{\small The red arrows represent ${\bf v}$ (normalized to unity), defined from~\eqref{vrvt} with $H_+=\sqrt{1-2/r}/(r\sin\theta)$, on a portion of the $(r,\theta)$ plane for the Schwarzschild BH with unit mass, in standard coordinates. The LR sits at $r=3$, $\theta=\pi/2$. Circulating the contour $C_1$ (or any  contour that encloses the LR) anti-clockwise, ${\bf v}$ winds once clockwise (follow the blue arrows $1\rightarrow 10$). Thus $w=-1$. By contrast, circulating the contour $C_2$ (or any contour that does not enclose the LR) anti-clockwise, ${\bf v}$ has no winding. Thus $w=0$. Observe two important properties that will be general. $(1)$ ${\bf v}$ becomes vertical at $\theta=0$ ($\theta=\pi)$ and downwards (upwards) directed; $(2)$ $v_r$ is positive (negative) as the horizon (asymptotic infinity) is approached. The signs are reversed for $H_-$.}
\label{fig1}
\end{center}
\end{figure}

Fig.~\ref{fig1} exhibits ${\bf v}$ for a Schwarzschild BH. It illustrates that $w=-1$ ($w=0$) for any contour that encloses (does not enclose) the Schwarzschild LR. In general, if $C$ encloses a single saddle point (maximum/minimum) of the potential $H_\pm(r,\theta)$, then $w=-1$ ($w=+1$). A LR with $w=-1$  ($w=+1$) is dubbed \textit{standard} (\textit{exotic}). LRs in Schwarzschild/Kerr are standard. Furthermore, for any $C$, the total $w$ is the sum of the individual LR charges within $C$. In particular, if there are no LRs within $C$, then $w=0$.

Our task is to show that the total LR topological charge in the region outside a BH (under the assumptions stated above) is $w=-1$, regardless of choosing $H_+$ or $H_-$. This implies that at least one standard LR must exist within that region, for each rotation sense of the BH, and establishes the theorem. To achieve this we must select an appropriate contour.

%
%

{\bf The contour.}
For our generic $(\mathcal{M},g)_{\rm BH}$, we define a contour $C$ that encompasses a sub-region $\mathcal{I}$ of the orthogonal 2-space exterior to the horizon. Then, taking appropriate limits, $\mathcal{I}$ becomes the full exterior region.

\begin{figure}[ht!]
\begin{center}
\includegraphics[width=0.4\textwidth]{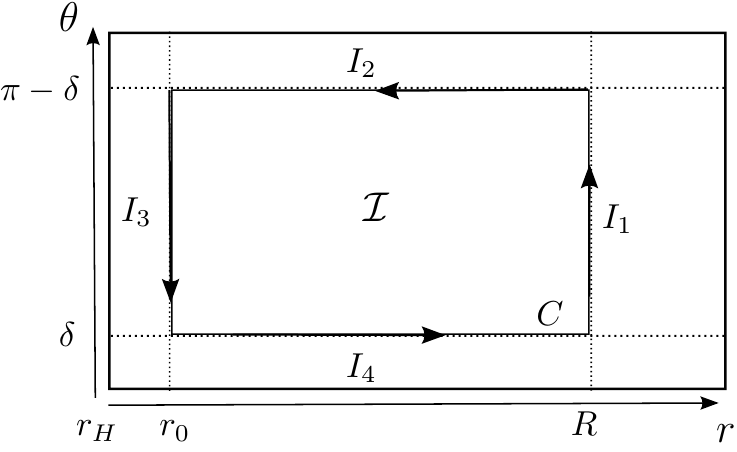}
\caption{\small Representation of the contour $C$ (which encloses $\mathcal{I}$) on the $(r,\theta)$ plane. The curve $C$ has positive orientation and it is composed by four line segments.}
\label{fig2}
\end{center}
\end{figure}

The region $\mathcal{I}$ is shown in Fig.~\ref{fig2} and it is defined as $r_0\leqslant r \leqslant R$ and $\delta\leqslant \theta \leqslant \pi-\delta$. The constants $\{r_0, R, \delta\}$ are such that $r_H<r_0\ll R$ and $0<\delta\ll 1$.

$\mathcal{I}$ is the region enclosed by the curve $C$ (see Fig.~\ref{fig2}), which is defined as the union of four line segments: $\{r=R, \, \delta\leqslant \theta \leqslant \pi-\delta\}\,\cup\,\{\theta=\pi-\delta, \,r_0\leqslant r \leqslant R\}\,\cup \{r=r_0, \, \delta\leqslant \theta \leqslant \pi-\delta\} \,\cup\, \{\theta=\delta,\, r_0\leqslant r \leqslant R\}$.

The topological charge of $\mathcal{I}$ is computed from~\eqref{tc}, decomposed as $2\pi w_{\mathcal{I}}=I_1+I_3 +I_2+I_4$, where:
\begin{equation}
I_1=\left[\int_\delta^{\pi-\delta}\frac{d\Omega}{d\theta}\,d\theta\right]_{r=R} \ , \ \ \  
I_2=\left[\int_R^{r_0}\frac{d\Omega}{dr}\,dr\right]_{\theta=\pi-\delta} \ , 
\end{equation}
\begin{equation}
I_3=\left[\int^\delta_{\pi-\delta}\frac{d\Omega}{d\theta}\,d\theta\right]_{r=r_0} \ , \ \ \
I_4=\left[\int^R_{r_0}\frac{d\Omega}{dr}\,dr\right]_{ \theta=\delta}\ .
\end{equation}

To obtain the total topological charge of the exterior region, we take first $\delta\to 0$ (axis limit), and only then $r_0\to r_H$ (horizon limit) and $R\to +\infty$ (asymptotic limit):
\be
w=\lim_{R\to+\infty}\lim_{r_0\to r_H}\left(\lim_{\delta\to  0}    w_{\mathcal{I}}\right) \ .
\label{totaltc}
\ee
These limits must be taken with care, as we now discuss.
  
{\bf Axis limit.}
The axis is the set of points for which $g_{\varphi\varphi}=\eta\cdot\eta\,=0=\eta\cdot\xi=g_{t\varphi}\,$. 
To approach the axis, introduce a local coordinate $\rho$, defined as $\rho\equiv \sqrt{g_{\varphi\varphi}}$ (recall $g_{\varphi\varphi}>0$ outside $\mathcal{H}$). 
Clearly, $d\rho/d\theta$ is positive (negative) as $\theta$ $\to$ $0$ ($\theta \to\pi$).
Then, consider a small $\rho$ expansion close to the axis:
\begin{align}
g_{\varphi\varphi}= \rho^2\ ,\qquad\qquad& g_{t\varphi}\simeq b_o\,\rho^n + \mathcal{O}(\rho^{n+1})\ ,\\
g_{tt}\simeq g_{tt}^{_0} +\mathcal{O}(\rho) \ ,\qquad\qquad& g_{\rho\rho}\simeq g_{\rho\rho}^{_0} +\mathcal{O}(\rho) \ ,
\end{align}
where $n\in\mathbb{N}$ and some constants were introduced. By assuming $C^2$-smoothness and regularity ($e.g.$ a non-diverging Ricci scalar) close to the axis $g_{\varphi\varphi}$ cannot go to zero faster than $g_{t\varphi}$ in the axis limit (see Appendix B and~\cite{carot2000some}). Then $2\leqslant n$ and $\rho^{2n}\ll \rho^2$. It follows from the definition of $D$ that $\sqrt{D}\simeq \rho\sqrt{-g_{tt}^{_0}}$.
Hence, from~\eqref{poth}:
\begin{equation}
H_\pm \simeq  \pm\frac{\sqrt{-g_{tt}^{_0}}}{\rho}\ .
\end{equation}
One can now estimate ${\bf v}$ from~\eqref{vrvt}. In particular, using $g_{\rho\rho}\,d\rho^2\simeq g_{\theta\theta}\,d\theta^2$ at zeroth order in $\rho$:
\be
v_\theta\simeq \, \textrm{sign}\!\left(\frac{d\rho}{d\theta}\right) \frac{\partial_\rho H_\pm}{\sqrt{g_{\rho\rho}}} \sim \,\,\mp\textrm{sign}\!\left(\frac{d\rho}{d\theta}\right) \frac{1}{\rho^2} \ .
\ee

Since $v_\theta\sim \rho^{-2}$ and $v_r\sim \rho^{-1}$, then $ v_\theta^2 \gg v_r^2$, and so $v\simeq |v_\theta|$. 
Hence as $\rho\to 0$ one obtains 
$v_\theta/v\to \mp\,\textrm{sign}\!\left(d\rho/d\theta\right)$.
 Consequently, 
\be \Omega=\left.\arcsin\left(\frac{v_\theta}{v}\right)\right|_{0,\pi} \to 
  \begin{cases} 
   \pm \pi/2 & \text{for } \theta\to\pi \\
   \mp \pi/2       & \text{for } \theta\to 0.
  \end{cases}\ee

The axis limit is $\lim_{\delta\to 0} C$, which implies $\rho\to 0$ along the integration paths of $\{I_2,I_4\}$. Thus, the bottom line is that $\Omega$ becomes constant along the integration path. Consequently, the contribution of $\{I_2,I_4\}$ to $w$ \textit{vanishes} as $\delta\to 0$.

This result can be interpreted as follows. In a generic BH spacetime, the arrows analogue to those in Fig.~\ref{fig1} become vertical along $\{I_2,I_4\}$ as $\delta\to 0$, directed upwards (downwards) at $\theta=\pi$ and downwards (upwards) at $\theta =0$, for $H_+$ ($H_-$). Hence, the integration along these paths does  not contribute to the winding of ${\bf v}$, as $C$ is circulated.

{\bf Horizon limit.}
To address the horizon  limit ($r_0\to r_H$) we observe that, as discussed in~\cite{Medved:2004tp}, the metric near the Killing horizon of a generic stationary and axially-symmetric BH is fairly constrained if we require regularity ($e.g.$ finite Ricci scalar at horizon). If the BH is not extremal $(\kappa\neq 0)$, we can set a local radial coordinate $x$ such that $g_{xx}=1$ and $\left.x\right|_{\mathcal{H}}=0$ at the horizon. We also define $N=\sqrt{D/g_{\varphi\varphi}}$ and $\omega=-g_{t\varphi}/g_{\varphi\varphi}$, which yields $H_\pm=\omega \pm N/\sqrt{g_{\varphi\varphi}}$. Then, near the horizon~\cite{Medved:2004tp}: 
\begin{equation} 
\omega\simeq \omega_{_H} +  \mathcal{O}(x^2) \ , \ 
N \simeq \kappa\,x + \mathcal{O}(x^3)\ , \ 
g_{\varphi\varphi}\simeq g_{\varphi\varphi}^{_H} + \mathcal{O}(x^2)\,.
\end{equation}
This leads to:
\be
\partial_x H_\pm \,\simeq\, \pm\frac{\kappa}{\sqrt{g_{\varphi\varphi}^{_H}}} + \mathcal{O}(x) \ .\ee
Since $\frac{1}{\sqrt{g_{xx}}}\frac{\partial}{\partial x} = \frac{1}{\sqrt{g_{rr}}}\frac{\partial}{\partial r}$, then near the horizon $(x\simeq 0)$:
\begin{equation}
v_r=\frac{\partial_rH_\pm}{\sqrt{g_{rr}}}\simeq \pm \frac{\kappa}{\sqrt{g_{\varphi\varphi}^{_H}}}\ .
\end{equation}
Thus, we have the following horizon limit:
\begin{equation}
\left.\textrm{sign}(v_r)\right|_{\mathcal{H}}=\pm 1 \ .
\end{equation}
This is sufficient for our purpose. It means that ${\bf v}$ has a positive (negative) radial component along $I_3$ for $H_+$ ($H_-$), in the horizon limit. By continuity,  along $I_3$ ${\bf v}$ interpolates between an upwards (downwards) directed ${\bf v}$ at the intersection with $I_2$ - see Fig.~\ref{fig1} - and a downwards (upwards) directed ${\bf v}$ at the intersection with $I_4$, for $H_+$ ($H_-$). Its positive (negative) radial component along $I_3$,  means ${\bf v}$ winds in the negative, $i.e.$ clockwise, direction along $I_3$, producing half of a full winding. Thus
\be
\Omega_{\theta=0}^\mathcal{H}-\Omega_{\theta=\pi}^\mathcal{H}=-\pi \ .
\label{fh}
\ee

{\bf Asymptotic limit.}
Finally consider the limit $R\to \infty$ (integration path of $I_1$). One reaches flat spacetime in standard spherical coordinates, yielding:
\begin{equation}v_r\simeq \mp\,\frac{1}{r^2\sin\theta}\quad \implies \quad \left.\textrm{sign}(v_r)\right|_{\infty}=\mp 1\ .
\label{vinf}
\end{equation}
Again, this information suffices: ${\bf v}$ has a negative (positive) radial component along $I_1$ for $H_+$ ($H_-$). A similar reasoning to that discussed above for the horizon limit, means ${\bf v}$ winds in the negative ($i.e.$ clockwise) direction along $I_1$, when $C$ is circulated in the positive ($i.e.$ counter-clockwise) direction, producing another half of a full winding. This means
 \be
\Omega_{\theta=\pi}^\infty-\Omega_{\theta=0}^\infty=-\pi \ .
\label{sh}
\ee

{\bf Total topological charge in the exterior region.}
The limits discussed above imply that the topological charge within $\mathcal{I}$, computed from~\eqref{totaltc} is $w=-1$, corresponding to a full winding of ${\bf v}$ in the negative sense as the contour delimiting $\mathcal{I}$ is circulated in the positive sense. Indeed,~\eqref{totaltc} reduces to
\be
w=\frac{1}{2\pi}\left[\int_0^{\pi}d\Omega\right]_{r=\infty}+\frac{1}{2\pi}\left[\int^0_{\pi}d\Omega\right]_{r=r_H} \ ,
\ee
or
\be 
w = \frac{1}{2\pi}\left(\Omega_\pi^\infty-\Omega_0^\infty + \Omega_0^\mathcal{H} - \Omega_\pi^\mathcal{H}\right)=-1 \ ,
\label{result}
\ee
where~\eqref{fh} and~\eqref{sh} were used in the last equality. This holds for both $H_\pm$ and means that there exists at least one standard LR (saddle point of $H_\pm$) for each rotation sense, in the exterior of the BH. Thus, the theorem is proved.

\begin{figure}[h!]
\begin{center}
\includegraphics[width=0.45\textwidth]{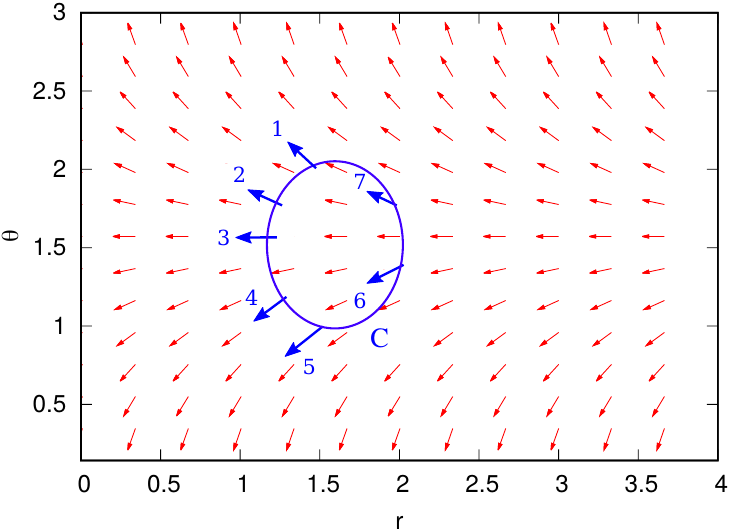}
\includegraphics[width=0.45\textwidth]{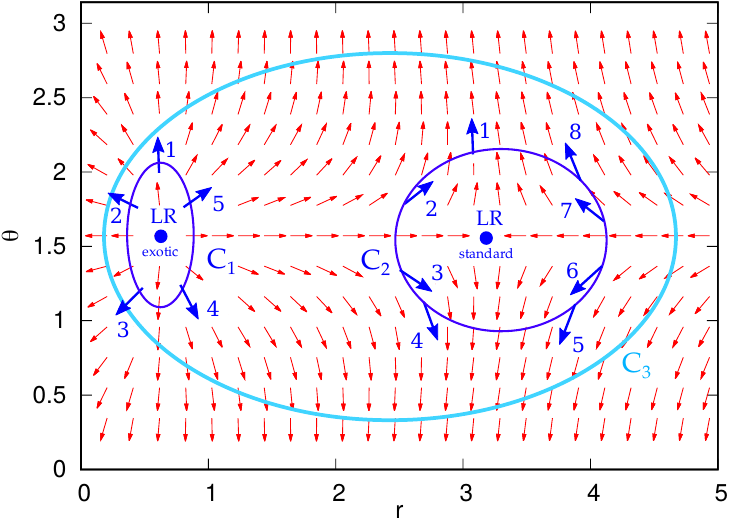}
\caption{\small {\it Top:} the red arrows represent ${\bf v}$, defined from~\eqref{vrvt} with $H_+=1/(r\sin\theta)$, on a portion of the $(r,\theta)$ plane for flat spacetime, in standard coordinates. There are no LRs. Observe the key difference with respect to Fig.~\ref{fig1}. Here,  $v_r$ is negative (positive for $H_-$) as the left boundary of the domain is approached, which is now a regular origin at $r=0$, rather than a horizon. Any contour $C$ will have $w=0$. {\it Bottom:} ${\bf v}$ (defined from $H_+$) for a horizonless ultra-compact object (rotating boson star). There are two LRs with opposite topological charge. By circulating the contour $C_1$ ($C_2$) anti-clockwise, ${\bf v}$ winds once in the positive (negative) sense (follow the numbered blue arrows). By contrast, after a full circulation along $C_3$, which encompasses both LRs, ${\bf v}$ winds up zero times $(w=0)$.}
\label{fig3}
\end{center}
\end{figure}

{\bf Absence of a horizon.}
To understand the key importance of the horizon $\mathcal{H}$, consider the potential $H_{\pm}$ for flat spacetime - see top row of Fig.~\ref{fig3}.
As expected the essential difference occurs near the left edge of Fig.~\ref{fig3} (top row). The absence of a horizon means ${\bf v}$ keeps flowing towards the left in the whole domain, $i.e.$ $v_r=-1/(r^2\sin\theta)<0$, with the sole exception of the axis limit, where it becomes vertical, since $v_\theta/v_r=\cot\theta\rightarrow \pm \infty$ at $\theta=0,\pi$, respectively. It is the presence of a horizon that introduces the $v_r>0$ boundary behaviour at the left boundary of the $(r,\theta)$ domain. As our theorem shows, this new boundary behaviour \textit{must} introduce (at least) one LR for each rotation sense.  

For flat spacetime, $w=0$ for any contour, and, in particular, one that encloses the full $(r,\theta)$ plane, as it is clear from Fig.~\ref{fig3} (top row). This is true, in fact, as long as the behaviour at all boundaries is kept, even for a curved spacetime. Thus, smoothness at the origin and at the axis, together with asymptotic flatness guarantees that the total topological charge will remain zero $w=0$, for any axi-symmetric, stationary spacetime, which is smoothly deformable into flat spacetime (and circular). Nonetheless, in such generic  smooth horizonless spacetime ${\bf v}$ may be locally deformed in the bulk so that LRs emerge. LRs \textit{do not} require a horizon. The individual LR charges, however, must add up to zero. In particular, for each standard LR (a saddle point of $H_\pm$, thus with $w=-1$) there must be a non-standard LR (maximum or minimum, thus with $w=+1$). This is the theorem in~\cite{Cunha:2017qtt}. Moreover, if the null energy condition is obeyed, the non-standard LRs must be stable. Thus, horizonless, asymptotically flat spacetimes with LRs must have a stable LR as long as they are a smooth deformation from flat spacetime, like those originating from an incomplete gravitational collapse~\cite{Cunha:2017qtt}. This is illustrated in Fig.~\ref{fig3} (bottom) where ${\bf v}$ is exhibited for an ultra-compact rotating boson star, a horizonless object in Einstein-Klein-Gordon theory~\cite{Ruffini:1969qy,Cunha:2018acu}. Observe that $w=\{+1,-1,0\}$ respectively for the contours $\{C_1,C_2,C_3\}$.

{\bf Discussion.}
Our theorem puts on a firm ground the hitherto unproved expectation that \textit{generic} equilibrium BHs must have one standard LR (for each rotation sense), (see also~\cite{Hod:2017xkz,CP}). In addition, it suggests possible ways to circumvent this result. For instance, by dropping: $(i)$ the 
circularity of the metric. Spacetime circularity holds in vacuum GR BHs but there are reasonable scenarios wherein it can be violated ($e.g.$ toroidal magnetic fields~\cite{gourgoulhon1993noncircular}). There is no fundamental reason for circularity to hold for astrophysical BHs; $(ii)$
asymptotic flatness. Changing the asymptotic behaviour of the spacetime may change the boundary behaviour~\eqref{vinf} and hence the whole result. The powerful tool of contour integration and topological LR charge may help understand more general situations.
It seems possible to tackle extremal BHs or non-spherical ($e.g.$ toroidal) horizons in a similar way. Astrophysically one does not expect extremal BHs, which are thus not the focus of this work. Moreover, recall that for extremal Kerr, the Boyer-Lindquist radial coordinate of the co-rotating LR coincides with that of the horizon. This is a coordinate artifact, but it suggests that the extremal BHs analysis introduces subtleties.

Finally, some of our  assumptions are implied if one focuses on GR with physical matter. For instance, assuming a GR stationary BH spacetime that is asymptotically flat and regularly predictable, with matter satisfying the dominant energy condition, then by Hawking's theorem~\cite{Hawking:1971vc} the cross-section of the event horizon has to be \textit{topologically spherical} ($S^2$), and the event horizon is a \textit{Killing horizon}. By further assuming that the spacetime is {analytic}, non-static and with the ergo-sphere intersecting the horizon, the spacetime is then required to be {\it axially symmetric} by Hawking's rigidity theorem.


{\bf Acknowledgements.}
We thank E. Berti, J. Joudioux, J. Nat\'ario, C. Paganini E. Radu, M. Rodriguez for discussions. This work is supported by the Center for Research and Development in Mathematics and Applications (CIDMA) through the Portuguese Foundation for Science and Technology (FCT - Fundacao para a Ci\^encia e a Tecnologia), references UIDB/04106/2020 and UIDP/04106/2020 and by the projects PTDC/FIS-OUT/28407/2017 and CERN/FIS-PAR/0027/2019. This work has further been supported by  the  European  Union's  Horizon  2020  research  and  innovation  (RISE) programme H2020-MSCA-RISE-2017 Grant No.~FunFiCO-777740. P. C. is supported by the Max Planck Gesellschaft through the Gravitation and Black Hole Theory Independent Research Group. The authors acknowledge networking support by the COST Action CA16104.

\bibliography{letterbhlr}


{\bf Appendix A.}
The sign of $H_\pm$ at a LR determines the LR's rotation sense, $i.e.$ $d\varphi/dt$ - see the discussion in~\cite{Cunha:2016bjh} in terms of the inverse function $h_\mp=H_\pm^{-1}$.  
If the LR is {\it outside} the ergoregion $(g_{tt}<0)$, then $D>g_{t\varphi}^2$. Thus,  $\textrm{sign}\!\left(H_\pm\right)=\pm 1$. For, say, negative BH angular momentum, which means $g_{t\varphi}>0$, a critical point of the potential $H_+$ ($H_-)$ outside the ergoregion is a counter-rotating (co-rotating) LR. For $g_{t\varphi}<0$ the roles of the potentials is swapped. 


{\it Inside} the ergoregion $g_{tt}>0$. Thus $|g_{t\varphi}|>\sqrt{D}$, and so $\textrm{sign}\!\left(H_\pm\right)=\textrm{sign}(-g_{t\varphi})$. Although the ergoregion may be composed by several disconnected spacetime regions, $g_{t\varphi}$ has a constant sign within each connected section of the ergoregion; otherwise there would be a metric signature change.

To be concrete, assume $g_{t\varphi}>0$. Then, there must be a sign change for $H_+$: it is negative (positive) when inside (outside) the ergoregion, and zero at the boundary. Outside it describes a counter-rotating LR. But there can be no counter-rotating LRs inside the ergoregion. By contrast, $H_-$ does not change sign. 

Let $\mathcal{B}$ be the outermost boundary of a connected ergoregion. $\mathcal{B}$ can have two topologies: i) spherical and enclosing the horizon ($e.g.$ in Kerr); or ii) toroidal ($e.g.$ in rotating boson stars). Consider case i) first. One can redefine the radial coordinate such that $\mathcal{B}$ is at constant $r$. Then $v_r>0$ at $\mathcal{B}$ due to the way $H_+$ changes sign at $\mathcal{B}$. The boundary behaviour of ${\bf v}$ at $\mathcal{B}$ is thus similar to that at $\mathcal{H}$ in Fig.~\ref{fig1}. It follows that $w=-1$ outside of $\mathcal{B}$ for $H_+$. Hence there is still a counter-rotating LR outside the ergoregion. 
For case ii), $\mathcal{B}$ is topologically a circle in the $(r,\theta)$ plane, with {$\bf v$} defined from $H_+$ always pointing outwards, due to the way $H_+$ changes sign. Thus, similarly to contour $C_1$ in Fig.~\ref{fig3}, ~$\mathcal{B}$ has $w=+1$. This means  an ergo-torus is accompanied by an exotic LR. Consequently, there must be a $w=-2$ contribution  from outside $\mathcal{B}$, and thus two counter-rotating LRs outside $\mathcal{B}$.

\vspace*{0.2cm}
{\bf Appendix B.}
Let us show that $g_{t\varphi}$ has to fall off as fast (or faster) than $g_{\varphi\varphi}$ when approaching the axis.
The following form of the metric will be used $ds^2=g_{tt}(\rho,z)dt^2 + 2g_{t\varphi}(\rho,z)\,dtd\varphi + \rho^2 d\varphi^2 + g_{\rho\rho}(\rho,z)d\rho^2 + g_{zz}(\rho,z)dz^2$, 
where both $\rho\equiv \sqrt{g_{\varphi\varphi}}$ and $g_{t\varphi}$ vanish at the axis ($\rho=0$).
Since the metric is $C^2$-smooth, we take the following expansion close to the axis:
\begin{align} 
&g_{t\varphi}\simeq g_{t\varphi}^{(1)}(z)\,\rho + g_{t\varphi}^{(2)}(z)\,\rho^2+ \mathcal{O}(\rho^{3}) \ , \nonumber \\
&g_{tt}\simeq -1 + \mathcal{O}(\rho)\ ,\ \ \
g_{\rho\rho}\simeq 1 + \mathcal{O}(\rho)\ , \ \ \ g_{zz}\simeq 1 + \mathcal{O}(\rho)\ , \nonumber \end{align}
where the zeroth order value of $\{g_{tt},g_{\rho\rho},g_{zz}\}$ is unity (in modulus) by redefinition of the respective coordinates.
One then obtains for the Ricci scalar:
\[R\simeq \frac{1}{\rho^2}\left(\frac{a^2(1+a^2) + \mathcal{O}(\rho)}{2(1+a^2)^2 + \mathcal{O}(\rho)}\right)\simeq \left(\frac{a^2}{1+a^2}\right)\frac{1}{2\rho^2} + \mathcal{O}\left(\frac{1}{\rho}\right),\]
where $a\equiv g_{t\varphi}^{(1)}(z)$. Thus, a \textit{necessary} condition for the curvature invariant $R$ to be finite as $\rho\to 0$ is for $a=0$. Hence $g_{t\varphi}$ has to go to zero as $\rho^2$ or faster, $i.e.$ as $\rho^n$, with $n\geqslant 2$, $n\in\mathbb{N}$.

 
\end{document}